\documentstyle[twoside,fleqn,epsfig,espcrc2]{article}

\newcommand{\AmS}{{\protect\the\textfont2
  A\kern-.1667em\lower.5ex\hbox{M}\kern-.125emS}}

\title{Q-balls in Underground Experiments}

\author{M. Ouchrif~\address{Dipartimento di Fisica dell'Universit\`a 
	di Bologna\\ 
        and INFN Sezione di Bologna,\\ 
        Viale Berti  Pichat 6/2, I-40127, Italy.\\ 
	and\\ 
	Universit\'e Mohamed Premier, Facult\'e des Sciences,\\
	D\'epartement de Physique, L.P.T.P.\\
	B.P: 524 ~~60000~Oujda,~~Morocco }        
\thanks{e-mail: ouchrif@bo.infn.it}}
       
\begin{document}

\begin{abstract}
In this paper we present some features of Q-balls and we
discuss their interactions with matter, and their
energy losses in the
Earth, for a large range of velocities. These calculations are used to compute
the fractional geometrical acceptance of the MACRO detector. Furthermore a
systematic
analysis of the energy losses of Q-balls in three types of detectors is
investigated. More specifically we have computed the light yield in liquid
scintillators, the ionization in streamer tubes and the Restricted Energy Loss
 in the $CR39$ nuclear track detectors.

\end{abstract}

\maketitle

\section{INTRODUCTION}
Dark Matter (DM) is one of the most intriguing problems in particle physics
and cosmology. Several types of stable particles
hypothesized in theories beyond the Standard Model of particle physics
have been considered as candidates for DM. One example of such particles
is the lightest supersymmetric particle (LSP) coming
from a supersymmetric extension of the Standard Model
\cite{TDLEE92}.
 
In theories where scalar fields,
carry a conserved global quantum number,
$Q$, there may exist non-topological solitons which are stabilized by global
charge conservation. These particles are spherically symmetric and for
large values of $Q$  their masses and volumes grow linearly with $Q$. Thus they
act like homogenous balls of ordinary matter, with $Q$ playing the role of
particle number; Coleman called this type of matter {\em Q-balls}
\cite{Coleman85}.

The conditions for the existence of absolutely stable Q-balls are
satisfied in supersymmetric theories with low energy supersymmetry breaking. 
According to \cite{Kusenko98A}
abelian non-topological solitons with Baryon and/or Lepton
quantum numbers naturally appear in the spectrum of the Minimal
Supersymmetric Standard Model. The role of conserved quantum number
is played by the baryon number.
The same reasoning applies to sleptons for the lepton number and also to
scalar Higgs particles.
Q-balls can thus be considered like
coherent states  of squarks, sleptons and Higgs fields.
Under certain assumptions about the internal self interaction of
these particles
and field the Q-balls are absolutely stable \cite{Kusenko98A}.

In this note we recall the main physical and astrophysical properties
of these particles,
the interaction of
Q-balls with matter and their energy losses in matter,
and the possibility of traversing the Earth to reach the
MACRO detector \cite{Ouchrif98}. We neglect the possibility of : \\
$(i)$ electromagnetic radiation emitted by Q-balls of high $\beta$. \\
$(ii)$ strong interaction of Q-balls in the upper atmosphere capable of 
destroying the Q-ball \cite{TDLEE92}; this last point deserves further 
investigations.

\section{PROPERTIES OF Q-BALLS}
Q-balls could have
been produced in the Early Universe, and could  contribute to the DM.
Several mechanisms could have lead to the formation of Q-balls in the
Early Universe. They  may have been
created in the course of a phase transition, which is sometime  called
``solitogenesis'', or they could  have
been produced via fusion in processes reminiscent of
the big bang
nucleosynthesis,  wich have been called
`` solitosynthesis'';  small Q-balls can be
pair-produced in high energy collisions  \cite{Kusenko97}.

The astrophysical
consequenses of Q-balls in many ways resemble those of
strange quark matter, ``nuclearites"; the peculiarity of Q-balls is that their
mass grows as $Q^{3/4}$, while for nuclearites the mass grows linearly with 
baryon number \cite{Witten84}.

The Q-ball mass and size are
related to its baryon number \cite{Kusenko98B}.
For a supersymetic potential $U(\phi) \sim M_{S}^4=constant$ for
large scalar $\phi$, the Q-ball mass $M$ and radius $R$ are given by 
\cite{Kusenko98B}.

\begin{equation}
M=\frac{4 \pi \sqrt{2}}{3}M_{S}~Q^{3/4}
\end{equation}

\begin{equation}
R=\frac{1}{\sqrt{2}}M_{S}^{-1}~Q^{1/4}
\end{equation}
The parameter $M_{S}$ is the energy scale of the SUSY breaking symmetry.
A stability condition was found in ref. \cite{Kusenko98B}:
the Q-ball mass $M$
is related to the nucleon mass $M_{N}$ by

\begin{equation}
M \leq Q~M_{N}
\end{equation}
From Eq. 1 and Eq. 3 one has the stability constraint \cite{Kusenko98B}:
\begin{equation}
Q \geq 1.6 \times 10^{15}\left (\frac{M_{S}}{1~TeV}\right)^4
\end{equation}

In Fig. 1 the allowed region for stable Q-balls is indicated.

\begin{figure}
        \mbox{ \epsfysize=8.2cm
            \epsffile{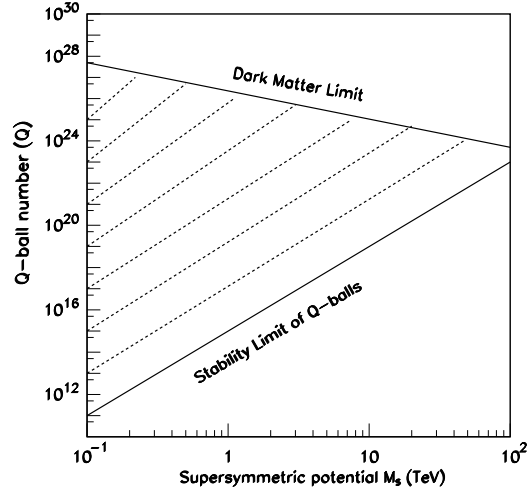}}
\vspace{-1cm}
\caption{Q-ball number versus the supersymmetry energy scale
$M_S$ for Q-balls. The shaded allowed
region is delimited by the Q-ball stability limit (Eq. 4) and by the
DM Limit, Eq. 6.}
\label{fig:largenenough}
\end{figure}

Q-balls are expected to concentrate in galactic
halos and to move with the typical galactic velocity
$v = \beta c \sim 10^{-3}c$.
Assuming that Q-balls constitute
the cold galactic dark matter with $\rho_{DM} \sim 0.3~GeV/cm^{3}$,
their number density is
\cite{Kusenko98B}
\begin{equation}
N_{Q} \sim \frac{\rho_{DM}}{M} \sim 5 \times 10^{-5}~Q^{-3/4}
\left(\frac{1~TeV}{M_{S}}\right)~cm^{-3}
\end{equation}

The corresponding flux is \cite{Kusenko98B}
\begin{equation}
\phi \sim \\ 10^{2}~Q^{-3/4}
\left(\frac{M_{S}}{1~TeV}\right)^{-1}~
cm^{-2}s^{-1}sr^{-1}
\end{equation}

If we assume that $\beta_{Q-ball} \sim 10^{-3}$, $\rho_{DM}
\sim 0.3~GeV/cm^{3}$, $M_{S} = 1~TeV$,  the Q-ball flux cannot be
greater that $\sim 4 \times \frac{10^{-19}}{M_{Q}} \;
cm^{-2} s^{-1} sr^{-1}$ ($M_{Q}$ in g). Q-balls are expected to be distributed
uniformly in our part of the galaxy and there should not be
enhancements in the solar system, as for example a cloud of Q-balls around
the sun.

Q-balls can be classified in two classes:
{\bf SECS} (Supersymmetric Electrically Charged Solitons) and {\bf SENS}
(Supersymmetric Electrically Neutral Solitons). The interaction of Q-balls with
matter and their detection differ drastically for SECS or SENS.

\section{INTERACTION WITH MATTER}

\subsection{Interaction with matter of Q-balls type SECS}
SECS are Q-balls with a net positive electric charge in the interior. The
charge of SECS originates from the unequal rate of absorption in the
condensate of quarks (squarks) and electrons (selectrons).
This
positive electric charge is
neutralized by a surrounding cloud of electrons. The positive charge
interacts via elastic
or quasi elastic collisions. The positive  electric charge can
vary from one to several tens;
the cross section is the Bohr cross
section for Q-ball hydrogen-interaction \cite{Kusenko98A}
\begin{equation}
\sigma = \pi a_{0}^{2}\sim 10^{-16}cm^2
\end{equation}
Where $a_{0}$ is the Bohr radius. The formula is valid  
for  $R \geq a_{0}$, which happens for $Q \geq 2.7 \times \left(
\frac{M_S}{1TeV}\right)^4$.

The main energy losses  \cite{Ouchrif98} of SECS passing throuth
matter with velocities in the
range $10^{-4}<\beta<10^{-2}$ is due to two contributions: the energy losses
due to $(i)$ the interaction of the SECS core
 with the nuclei (nuclear contribution)
and, $(ii)$ with the electrons of the traversed
medium  (electronic contribution).
The total energy loss is the sum of the two
contributions.

SECS could cause the catalysis of proton decay, but only if they are large
and have large velocities \cite{Pro}.
The possibility that SECS can cause the catalysis of proton decay  does not
concern our
range of interest for velocities, masses and radii of SECS.

{\bf Electronic losses of SECS:} The electronic contribution to the energy
loss of SECS is  calculated
with the following formula \cite{Ouchrif98}
\begin{equation}
\frac{dE}{dx} = \frac{8 \pi a_{0} e^{2}  \beta }{\alpha} \frac{Z_{1}^{7/6} 
N_{e}}{
(Z_{1}^{2/3} + Z_{2}^{2/3})^{3/2}}~~~~for~~Z_{1} \geq 1
\end{equation}
where $\alpha$ is the fine structure constant, $\beta = v/c$,
$Z_{1}$ is the positive core charge of SECS, $Z_{2}$ is the atomic number
of the medium and
$N_e$ is the density of electrons in the medium. Electronic losses dominate
for $\beta > 10^{-4}$.

{\bf Nuclear losses of SECS:} The nuclear contribution to the energy loss
of SECS is due to the interaction of the SECS positive core
with the nuclei of the medium and it is given by \cite{Ouchrif98}
\begin{equation}
\frac{dE}{dx} = \frac{\pi a^{2} \gamma N E }{\epsilon} S_{n}(\epsilon)
\end{equation}
where
\begin{equation}
S_{n}(\epsilon) \simeq \frac{0.56 Log(1.2\epsilon)}{1.2\epsilon -
(1.2\epsilon)^{-0.63}}~,~~~\epsilon = \frac{a M_{2} E}{Z_{1}Z_{2} e^{2}M_{1}}
\end{equation}
and
\begin{equation}
a= \frac{0.885 a_{0}}{(\sqrt{Z_{1}} + \sqrt{Z_{2}})^{2/3}}~,~~~~~~~~~~~~
\gamma= \frac{4 M_{2}}{M_{1}}
\end{equation}
$M_{1}=M$ is the mass of the incident Q-ball; $M_{2}$ is the mass of the
target nuclei; $Z_{1}e$ and $Z_{2}e$ are their electric charges; we assume
that $M_{1}> M_{2}$. Nuclear losses dominates for $\beta \leq 10^{-4}$.

{\bf The energy losses of SECS in the earth mantle and earth core:}

The energy losses of SECS in the earth mantle and earth core have been
computed for different $\beta$-regions and for different charges of the Q-ball
core, using the same general procedures used in the past for computing the
energy losses in the earth of magnetic monopoles and nuclearites
\cite{Ouchrif98}. 

In Fig. 2 is presented the energy losses of Q-balls type SECS in the Earth 
mantle.

\begin{figure}
        \mbox{ \epsfysize=8.5cm
        \epsffile{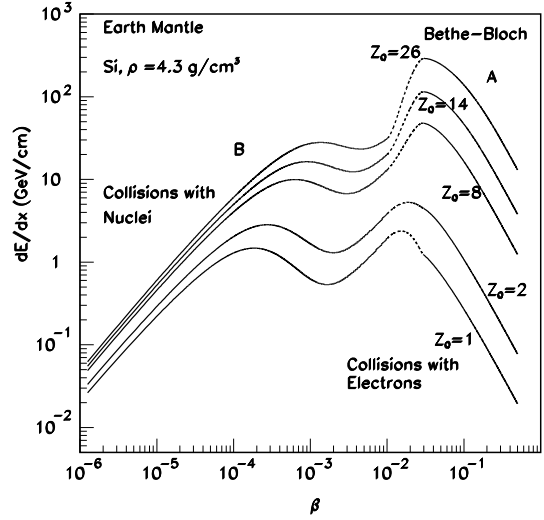}}
\vspace{-1cm}
\caption{Energy losses of SECS versus $\beta$ in the Earth Mantle. 
$Z$ is the electric charge of the Q-ball core.}
\label{yieldq}
\end{figure}

\subsection{Interaction of Q-balls type SENS}
The Q-ball interior of SENS is characterized by a
large Vacuum Expectation Value  (VEV) of certain squarks,
and may be sleptons and Higgs fields \cite{Kusenko98A}.
The $SU(3)_{c}$ symmetry is broken and deconfinement takes
place inside the Q-ball. If
a nucleon enters this region of deconfinement, it dissociates into three
quarks, some of
which may  then become absorbed in the condensate via the reaction \cite
{Kusenko98B}. 

\begin{equation}
qq \rightarrow \tilde{q} \tilde{q}
\end{equation}
In practice the reaction looks like
\begin{equation}
(Q) + Nucleon \rightarrow (Q+1) + pions
\end{equation}
and sometimes as
\begin{equation}
(Q) + Nucleon \rightarrow (Q+1) + Kaons
\end{equation}

If it is assumed that the energy released in (13) 
is the same as in typical
hadronic processes
(about 1~GeV per nucleon), this energy is  carried by 2 or 3 pions (or 2
kaons).
The cross section
for reactions (13) and (14)
 is determined by the Q-ball radius  $R$ \cite{Kusenko98A}
\begin{equation}
\sigma \sim 6 \times 10^{-34}Q^{1/2} \left(\frac{1~TeV}{M_S} \right)^{2}cm^2
\end{equation}

The corresponding mean free path $\lambda$ is
\begin{equation}
\lambda = \frac{1}{\sigma N}
\end{equation}
According to References [5-8] the
energy loss of SENS moving with velocities in the range
$10^{-4}<\beta <10^{-2}$ is constant and is given by
\begin{equation}
\frac{dE}{dx} \sim \frac{\zeta}{\lambda} = \zeta N 10^{-34}Q^{1/2} \left(\frac{1~TeV}{M_S} \right)^{2}cm^2
\end{equation}
where $N~is~the~number~of~atoms/cm^3$ and $\zeta$ is
the energy released
in the decay. The energy loss of SENS is due to the energy released from the
absorbed nucleon mass. SENS lose a very small fraction of their
kinetic energy and are able to traverse the Earth without attenuation for all
masses of our interest.

\section{ENERGY LOSSES IN DETECTORS}

\subsection{Light Yield of Q-balls type SECS} 
For SECS we distinguish two contributions to the Light Yield in scintillators:
the primary Light Yield  and the secondary Light Yield.

{\bf \em The primary Light Yield:} is due to the direct excitation and
ionization produced by the SECS in the medium. The energy losses in the MACRO
liquid scintillator is computed from the energy losses of protons in hydrogen
and carbon \cite{D97}
\begin{equation}
\left(\frac{dE}{dx} \right)_{SECS}= \frac{1}{14}\;\;[~2
\left(\frac{dE}{dx}\right)_{H} +
12 \left(\frac{dE}{dx} \right)_{C}] 
\end{equation}
\begin{equation}
= SP = \frac{SL \times SH}{SL + SH}
\end{equation}
where $SP$ is the stopping power of SECS, which  reduces to $SL$ at low
$\beta$ and to $SH$ at high $\beta$, so at very high $\beta$;
the $SP$ stopping power
coincides with the Bethe Bloch formula for electric
energy losses.
 
{\large \bf 1.} For $Q=1$ the energy losses of SECS in hydrogen and
carbon is computed
from \cite{Z77} adding an exponential factor due to the experimental data
\cite{F87}.

{\bf i)}~For $10^{-5}  <   \beta < 5 \times 10^{-3}$
we obtained the following formula
\begin{equation}
\left(\frac{dE}{dx} \right)_{SECS}
= C~[~1- exp( \frac{\beta}{
7 \times 10^{-4}})^{2}]~~~\frac{MeV}{cm}
\end{equation}
where $C=1.3 \times 10^{5} \beta $.\\
{\bf ii)}~For $5 \times 10^{-3}  <   \beta < 10^{-2}$ we used the following
formula \cite{F87}
\begin{equation}
SP = SP_{H} + SP_{C} = \left(\frac{dE}{dx} \right)_{SECS}
\end{equation}
where
\begin{equation}
SP_{H} = \frac{SL_{H} \times SH_{H}}{SL_{H} + SH_{H}}
\end{equation}

\begin{equation}
SP_{C} = \frac{SL_{C} \times SH_{C}}{SL_{C} + SH_{C}}
\end{equation}
and
\begin{equation}
SL = A_{1}~E^{0.45},~~~~~~~~~SH = A_{2}~Ln(1 + \frac{A_{3}}{E} + A_{4} E)
\end{equation}
where ($A_{i=1,4}$) are constants obtained from experimental data, and $E$ is
the energy of a proton with velocity $\beta$.

{\large \bf 2.} For SECS with $Q = Z_{1}e$ the energy losses for $ 10^{-5} <
\beta < 10^{-2}$
are given by
\cite{L61}
\begin{equation}
\left(\frac{dE}{dx} \right)_{SECS} = F(Z_{1},Z_{2})~[~1- exp(- \frac{\beta} 
{7 \times 10^{-2}})^2]
\end{equation}
where
\begin{equation}
F(Z_{1},Z_{2}) = 
\frac{ 8 \pi e^{2} a_{0} \beta}
{\alpha} \frac{Z_{1}^{7/6} N_{e}}
{(Z_{1}^{2/3} + Z_{2}^{2/3})^{3/2}} 
\end{equation}
where $Z_{2}$ is the atomic number of the target atom, $N_e$ the density
of electrons  and $\alpha$ is the fine structure constant.

The primary Light Yield of SECS is given by \cite{D97}
\begin{equation}
\left(\frac{dL}{dx} \right)_{SECS}
= A~[\frac{1}{1 + AB~\frac{dE}{dx}}]~\frac{dE}{dx}
\end{equation}
where $\frac{dE}{dx}$ is the energy loss of SECS;
$A$ and $B$ are parameters depending only on the velocity of SECS.

{\bf \em The secondary Light Yield:} we considered the elastic or quasielastic
recoil of hydrogen and carbon nuclei. The light yield $L_{p}$ from
a hydrogen or carbon nucleus of
given initial energy $E$ is computed as
\begin{equation}
L_{p}(E)=\int^{E}_{0}\frac{dL}{dx}(\epsilon)S^{-1}_{\mbox{tot}} \, d\epsilon
\end{equation}
where $S_{\mbox{tot}}$ is the sum of electronic and nuclear
energy losses. The  nuclear energy losses are given in ref. \cite{W77}.
The secondary light
yield is then
\begin{equation}
\left(\frac{dL}{dx}\right)_{\mbox{secondary}}=N\int^{T_{m}}_{0}L_{p}(T)
\frac{d\sigma}{dT} \, dT
\end{equation}
where $T_{m}$ is the maximum energy transferred and
 {\Large $\frac{d\sigma}{dT}$} is the
differential scattering cross section, given in ref. \cite{L77}.

In Fig. 3 is presented the light yield of SECS in MACRO liquid 
scintillator as function of the SECS velociy $\beta$. 

\begin{figure}
        \mbox{ \epsfysize=8.2cm
        \epsffile{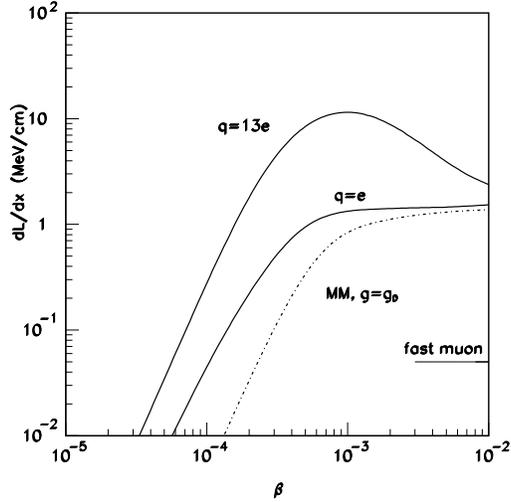}}
\vspace{-1cm}
\caption{Light Yield of SECS in the MACRO liquid scintillator as function of
the SECS velocity $\beta$; $q$ is the net positive electric charge of the
SECS core.}
\label{yieldq}
\end{figure}

\subsection{Energy losses of Q-balls type SECS in streamer tubes}
The composition of the gas in the  MACRO limited streamer tubes is 73\%
helium,  27\% n-pentane in volume \cite{Ouchrif98}.
The pressure is about one atmosphere and the resulting density
is low (in comparison with the density of the
other detectors): $ \rho_{gas}=0.856~\mbox{mg/cm}^{3}$.
 The energy losses of MMs in the streamer tubes have been discussed in
\cite{Ouchrif98}.

The ionization energy losses of SECS in the streamer tubes of
the MACRO experiment
for $10^{-3} < \beta < 10^{-2}$ was computed with the same procedure used
for scintillators, but using the density and the chemical composition of
streamer tubes.

For $Q =13e$ the energy losses are calculated from ref. \cite{L61}
but we have omitted the exponential factor which takes into account the
energy gap in organic scintillators.

The Drell effect does not  occurs because SECS are not magnetically charged.

The threshold for ionizing n-pentane occur at $\beta \sim 10^{-3}$.

\begin{figure}
\begin{center}
        \mbox{ \epsfysize=7.5cm
        \epsffile{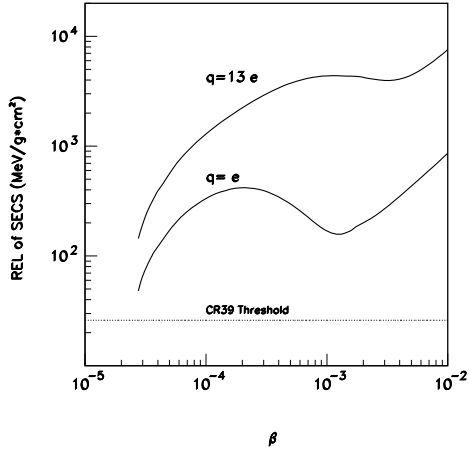}}
\vspace{-1cm}
\caption{Restricted Energy Losses of SECS as function of velocity in the
nuclear track detector $CR39$. The detection threshold for the MACRO CR39 is
also shown ref.~[5].}
\label{yieldq}
\end{center}
\end{figure}

\subsection{Restricted Energy Losses of SECS in the
Nuclear Track Detector CR39}
The quantity of interest for the CR39 nuclear track detector is the
Restricted Energy Loss (REL), that is, the energy deposited within $\sim$~
100~\AA~~diameter from the track.

The REL in CR39 has already been computed for MMs of $g=g_{D}$ and
$g=3g_{D}$ and for dyons with
$q = e$, $g=g_{D}$ \cite{R83}. We have
checked these calculations and extended them to other cases of interest
\cite{Ouchrif98}.

The~chemical~composition~of~CR39~is $\mbox{C}_{12}\mbox{H}_{18}\mbox{O}_{7}$~, 
and the
density is 1.31~$\mbox{g/cm}^3$.
For the computation of the REL only energy transfers to atoms above $12$~eV are
considered, because it is estimated that $12$~eV are necessary to break the
molecular bonds \cite{D97}.

At {\em low velocities} ($3 \times 10^{-5}<\beta<10^{-2}$) there are
two contributions to REL: the ionization and the atomic recoil
contributions.

{\em The ionization contribution} was computed
with Ziegler's fit to the experimental data \cite{Z77}.

{\em The atomic recoil contribution} to REL was calculated
using the  interaction potential between an atom and a SECS which is equal to
the electric potential \cite{W77} given by

\begin{equation}
V(r) = \frac{Z_{1} Z_{2} e^{2}}{r} \phi (r)
\end{equation}
where $r$ is the distance between the core of SECS and the target atom, $e$ is
the electric charge of the core,
$Z_{2}$ is the atomic number of the target atom. The
function $\phi (r)$ is the screening function given by \cite{W77}
\begin{equation}
\phi (r) = \sum_{1}^{3} C_{i}~exp[- \frac{b_{i} r}{a}]
\end{equation}
where $a$ is the screening length
\begin{equation}
a = 0.8853~\frac{a_0}{(Z_{1}^{\frac{1}{2}} + Z_{2}^{\frac{1}{2}})^
{\frac{2}{3}}}
\end{equation}
where $a_0$ is the Bohr radius; the coefficients  are restricted such that
\begin{equation}
\sum_{1}^{3} C_{i} =1
\end{equation}

Assuming the validity of this potential, we calculated the
relation between the scattering angle $\theta$ and the impact parameter b.
From this
relation, the differential cross section $\sigma (\theta )$ is readily obtained
as \cite{D97}
\begin{equation}
\sigma (\theta ) = -(db/d\theta)\cdot b/\sin \theta
\end{equation}
\hspace{0.6cm}
The relation between the transferred kinetic energy K and the scattering angle
$\theta$
is given by the relation
\begin{equation}
K = 4E_{\mbox{inc}}\sin^{2} (\theta/2)
\end{equation}
where $E_{\mbox{inc}}$ is the energy of the atom relative to the SECS
in the center of mass system. The Restricted
Energy Losses are finally obtained by
integrating the transferred energies as
\begin{equation}
-\frac{dE}{dx}  = N \int \sigma(K) \, dK
\end{equation}
where $N$ is the number density of atoms in the medium, $\sigma(K)$ is the
differential cross section as function of the transferred kinetic energy K.

In Fig. 4 is presented the restricted energy losses of the Q-ball type SECS in 
the nuclear solide track detector CR39.

\section{CONCLUSION}
Supersymmetric generalizations of the Standard Model allow for
stable non-topological solitons of Q-ball type which may be considered bags of
squarks and sleptons and thus have non-zero
baryon and lepton numbers, as well as the electric charge [1-3]. These solitons
can be produced in the Early Universe, can affect the nucleosynthesis
of light elements, and can
lead to a variety of other cosmological consequences.

In this paper, we computed the energy losses of Q-balls of type SENS and
SECS. Using these energy losses and a rough model of the earth's  composition
and density profiles, we have computed the geometrical acceptance of
the MACRO detector for Q-balls type SECS with $v=250~km/s$ as  function of
the Q-ball mass $M$. We have calculated the accessible region in the plane
(mass,  velocity) of SECS reaching MACRO from above and below.

We also presented a systematic analysis of the energy deposited in
scintillators, streamer tubes and $CR39$ nuclear track detectors by SECS in
forms useful for their detection. In particular we computed the light yield
in scintillators, the ionization in streamer tubes and the
REL in nuclear track detectors.

MACRO is sensitive to both SECS and SENS. A good upper flux limit may
be obtained at the level of few times $10^{-16}~~cm^{-2}s^{-1}sr^{-1}$.\\

{\bf Acknowledgements :}  
I gratefully acknowledge Prof. G. Gaicomelli for his continous 
disponibility and for very useful critical discussions.  
This work was supported by ICTP and INFN grants.


\begin{thebibliography}{9}
\bibitem{TDLEE92} T.D. Lee and Y. Pang, Phys. Rep. {\bf 221} (1992); \\
R. Ruffini and A. Bonnazola, Phys. Rev. {\bf 187} (1969) 1767;\\
J.D. Breit et al, Phys. Lett. {\bf B140} (1984) 329;\\
M. Colpi et al, Phys. Rev. Lett. {\bf 57} (1986) 2485.\\
\vspace{-0.5cm}
\bibitem{Coleman85} S. Coleman, Nucl. Phys. {\bf B 262} (1985) 293.\\
\vspace{-0.5cm}
\bibitem{Kusenko97} A. Kusenko, Phys. Lett. {\bf B 405} (1997) 108; 
Phys. Lett. {\bf B 404} (1997) 285; Phys. Lett. {\bf B 406} (1997) 26.\\
\vspace{-0.5cm}
\bibitem{Kusenko98A} A. Kusenko and M. Shaposhnikov, Phys. Lett.
{\bf B 417} (1998) 99;\\
A. Kusenko et al., Phys. Rev. Lett.
{\bf 80} (1998) 3185.\\
\vspace{-0.5cm}
\bibitem{Ouchrif98} J. Derkaoui et al, Astropart. Phys. {\bf 9} (1998) 173.\\
\vspace{-0.5cm}
\bibitem{Witten84} E. Witten, Phys. Rev. {\bf D 30} (1984) 272; \\
A. De R\`ujula and Glashow, Nature {\bf 312} (1984) 734; \\
E. Farhi and R.L. Jaffe, Phys. Rev. {\bf D 30} (1984) 2379.\\
\vspace{-0.5cm}
\bibitem{Kusenko98B} A. Kusenko, Phys. Rev. Lett. {\bf B 61} (1998) 2909;\\
J. Madsen, Phys. Lett. {\bf B 435} (1998) 125; 
Phys. Lett. {\bf B 246} (1990) 135;\\
A.V. Olinto '{\em The Physics of Strange Matter}', Proceedings of
'{\em  Relativistic Aspects of Nuclear Physics}', Rio de Janiero, Brazil
(1991);\\
J. Madsen, astro-ph/9809032 (1998).\\
\vspace{-0.5cm}
\bibitem{Pro} V.A. Rubakov, Rep. Prog. Phys. {\bf 51} (1988) 189.\\
\vspace{-0.5cm}
\bibitem{OUCHRIF} M. Ouchrif, '{\em Energy losses of Q-balls}' MACRO 
Int Memo, 7/99 (1999).\\
\vspace{-0.5cm}
\bibitem{Kolda} T. Gherghetta, C. Kolda and S.P. Martin, Nucl. Phys.
{\bf B468} (1996) 37;\\
G. Cleaver et al., hep-th/9711178 (1997).\\
\vspace{-0.5cm}
\bibitem{Z77} H. H. Andersen and J.F. Ziegler,
{\em Hydrogen stopping power and ranges in all elements}, Pergamon Press
(1977). \\
\vspace{-0.5cm}
\bibitem{F87} D. J. Ficenec et al., Phys. Rev. {\bf D36} (1987) 311. \\
\vspace{-0.5cm}
\bibitem{L61} J. Lindhard and M. Scharff, Phys. Rev. {\bf 124}
 (1961) 28.\\
\vspace{-0.5cm}
\bibitem{D97} J. Derkaoui al., Astropart. Phys. {\bf 10} (1999) 339.\\
\vspace{-0.5 cm}
\bibitem{W77} W. D. Wilson, L. G. Haggmark and J. P. Biersack,
Phys. Rev. {\bf B15}  (1977) 2458. \\
\vspace{-0.5cm}
\bibitem{L77} J. Lindhard et al,
K. Dan. Vidensk. Selsk. Mat.-Fys. Med. {\bf 33}  (1963) No. 14.\\
\vspace{-0.5cm}
\bibitem{R83} T. W. Ruijgrok, J. A. Tjon and T. T. Wu, Phys. Lett. {\bf 129B} (
S. Nakamura, Ph. D. Thesis, {\em Search for supermassive relics by large
area plastic track detectors}, UT-ICEPP-88-04, University of Tokyo (1988).\\




\end{thebibliography}
\end{document}